\documentclass[twocolumn,superscriptaddress,showpacs,prl]{revtex4}
\usepackage{graphicx}

\begin{document}

\title{Spin squeezing via quantum feedback}

\author{L. K. Thomsen}
\affiliation{Centre for Quantum Dynamics, School of Science,
Griffith University, Brisbane, Queensland 4111, Australia}

\author{S. Mancini}
\affiliation{Dipartimento di Fisica, Universit\`a di Camerino,
I-62032 Camerino, Italy} \affiliation{INFM, Dipartimento di
Fisica, Universit\`a di Milano, Via Celoria 16, I-20133 Milano,
Italy}

\author{H. M. Wiseman}
\affiliation{Centre for Quantum Dynamics, School of Science,
Griffith University, Brisbane, Queensland 4111, Australia}

\date{\today}

\begin{abstract}
We propose a quantum feedback scheme for producing
deterministically reproducible spin squeezing. The results of a
continuous nondemolition atom number measurement are fed back to
control the quantum state of the sample. For large samples and
strong cavity coupling, the squeezing parameter minimum scales
inversely with atom number, approaching the Heisenberg limit.
Furthermore, ceasing the measurement and feedback when this
minimum has been reached will leave the sample in the maximally
squeezed spin state.
\end{abstract}


\maketitle

\newcommand{\beq}{\begin{equation}}
\newcommand{\eeq}{\end{equation}}
\newcommand{\bqa}{\begin{eqnarray}}
\newcommand{\eqa}{\end{eqnarray}}
\newcommand{\nn}{\nonumber}
\newcommand{\nl}[1]{\nn \\ && {#1}\,}
\newcommand{\erf}[1]{Eq.~(\ref{#1})}
\newcommand{\dg}{^\dagger}
\newcommand{\smallfrac}[2]{\mbox{$\frac{#1}{#2}$}}
\newcommand{\half}{\smallfrac{1}{2}}
\newcommand{\sq}[1]{\left[{#1}\right]}
\newcommand{\cu}[1]{\left\{{#1}\right\}}
\newcommand{\ro}[1]{\left({#1}\right)}
\newcommand{\bra}[1]{\langle{#1}|}
\newcommand{\ket}[1]{|{#1}\rangle}
\newcommand{\ip}[2]{\langle{#1}|{#2}\rangle}
\newcommand{\an}[1]{\left\langle{#1}\right\rangle}
\newcommand{\ito}{It\^o }

Squeezed spin systems \cite{KitUed93} of atoms and ions have
attracted considerable attention in recent years due to the
potential for practical applications, such as in the fields of
quantum information \cite{QInf98} and high-precision spectroscopy
\cite{Spect}. Quantum correlations of squeezed spin states
outperform classical states in analogy with squeezed optical
fields. Moreover, squeezed spin sates are also multiparticle
entangled states \cite{Soretal01}. Recent proposals for their
generation include the absorption of squeezed light
\cite{sqzlight}, collisional interactions in Bose-Einstein
condensates \cite{Soretal01,collBEC}, and direct coupling to an
entangled state through intermediate states such as collective
motional modes for ions \cite{MolSor99} or molecular states for
atoms \cite{HelmYou01}.

Other proposals create spin squeezing via quantum nondemolition
(QND) measurements \cite{KuzBigMan98,KuzManBig00,Duanetal00}. A
striking recent achievement of QND measurements is the
entanglement of two macroscopic atomic samples \cite{JulKozPol01}.
These QND schemes produce conditional squeezed states that are
dependent on the measurement record. On the other hand,
unconditional squeezing would ensure that the state is
deterministically reproducible. M\o lmer \cite{Mol99} has shown
that alternating QND measurements and incoherent feedback can
produce sub-Poissonian number correlations. However, that work
does not treat the quantum effects of the measurement back action
or the feedback on the mean spin (which is assumed to be zero).
Hence it cannot predict the strength of the entanglement.

In this Rapid Communication, we suggest achieving spin squeezing
via feedback that is {\em coherent} and {\em continuous}. We
consider a continuous QND measurement of the total population
difference of an atomic sample. The results of the measurement,
which conditionally squeeze the atomic sample, are used to drive
the system into the desired, deterministic, squeezed spin state.
This involves amplitude modulation of a radio-frequency (rf)
magnetic field, where the feedback strength varies in time such
that the mean number difference is kept at zero.

An ensemble of $N$ two-level atoms can be described by a spin-$J$
system \cite{Dic54}, i.e., a collection of $2J=N$ spin-$\half$
particles. The collective spin operators are given by $J_{\alpha}
=\sum_{k=1}^{N}\half\sigma^{(k)}_{\alpha}(\alpha=x,y,z)$, where
$\sigma_{\alpha}^{(k)}$ are the Pauli operators for each particle.
Thus, $J_{z}$ represents half the total population difference.
Coherent spin states (CSS) have variances normal to the mean spin
direction equal to the standard quantum limit (SQL) of $J/2$.
Introducing quantum correlations among the atoms reduces the
variance below the SQL in one direction at the expense of the
other \cite{KitUed93}. Such squeezed spin states can be
characterized by the squeezing parameter \cite{Soretal01} \beq
\xi^{2}_{{\mathbf{n}}_{1}}= N(\Delta J_{{\mathbf{n}}_1})^2/
(\an{J_{{\mathbf{n}}_{2}}}^{2}+\an{J_{{\mathbf{n}}_{3}}}^{2}),
\label{xisq} \eeq where ${\mathbf n}_{i}(i=1,2,3)$ are orthogonal
unit vectors. Systems with $\xi^{2}_{\mathbf n}<1$ are spin
squeezed in the direction ${\mathbf n}$ and also have
multiparticle entanglement \cite{Soretal01}.

\begin{figure}
\includegraphics[width=0.45\textwidth]{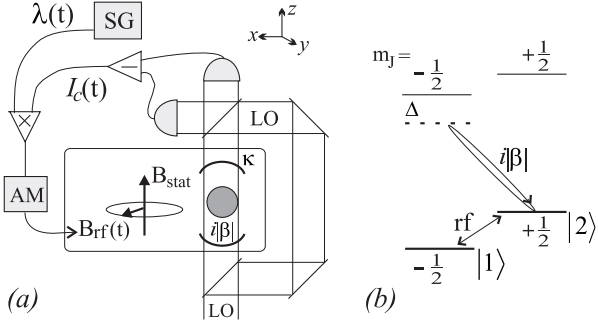}
\vspace{-0.3cm} \caption{\label{expt} $(a)$ Schematic experimental
configuration. A cavity field of amplitude $i|\beta|$ interacts
with the atomic sample. The current $I_{c}(t)$ from the homodyne
detection of the cavity output, damped at rate $\kappa$, is
combined with $\lambda(t)$ produced by a signal generator (SG).
The combined signal controls the amplitude (AM) of an rf magnetic
field that, together with a static field, drives $J_{y}$. $(b)$
Single atom diagram. The static B field lifts the degeneracy of
the magnetic sublevels. The far-detuned cavity field $i|\beta|$
monitors the collective population in state $\ket{2}$ (and hence
$J_{z}$). The rf driving field, applied perpendicularly to the
static field direction, induces magnetic dipole transitions
between $\ket{1}$ and $\ket{2}$ (thus driving $J_{y}$).}
\end{figure}

Let the internal states, $\ket{1}$ and $\ket{2}$, of each atom be
the degenerate magnetic sublevels of a $J=\half$ state, e.g., an
alkali ground state.  Each atom is prepared in an equal
superposition of the two internal states, thus giving a CSS of
length $J$ in the $x$-direction. The atomic sample is placed in a
strongly driven, heavily damped, optical cavity, as shown in
Fig.~\ref{expt}$(a)$. The cavity field is assumed to be far off
resonance with respect to transitions probing state $\ket{2}$, see
Fig.~\ref{expt}$(b)$. This dispersive interaction causes a phase
shift of the cavity field proportional to the number of atoms in
$\ket{2}$. Thus, the QND measurement of $J_{z}$ (since $N$ is
conserved) is effected by the homodyne detection of the light
exiting the cavity \cite{CorMil98}.

This interaction is defined by the Hamiltonian $\hbar\chi J_{z}
b\dg b$ where $b$,$b\dg$ are the cavity field operators and
$\chi=g^2/4\Delta$, with one-photon Rabi frequency $g$, and
optical detuning $\Delta$ \cite{CorMil98}. For strong coherent
driving we can use the semiclassical approximation $b\to
i|\beta|+b$, where now $b$ represents small quantum fluctuations
around the classical amplitude $i|\beta|$. The interaction is thus
\beq
H_{\rm int} = \hbar\chi|\beta|J_{z}(-ib+ib\dg), \label{Hint}
\eeq
where we have chosen an initial $J_{z}$ splitting of
$-\chi|\beta|^2$.

Following the procedure of Sec. VII in Ref.~\cite{WisMil94} we can
adiabatically eliminate the cavity dynamics if the cavity decay
rate $\kappa\gg\chi|\beta|J_{z}$, which requires
$\kappa\gg\chi|\beta|\sqrt{N}$ (since the initial $\Delta
J_{z}=\sqrt{J/2}$). The evolution of the atomic system due the
measurement is thus \beq \dot{\rho}=M{\cal D}[J_{z}]\rho,
\label{MEmeas} \eeq where $M =4\chi^{2}|\beta|^{2}/\kappa$ is the
measurement strength [equivalent to $2D$ in Eq.~(22) of
Ref.~\cite{CorMil98}], and ${\cal D}[r]\rho\equiv r\rho r\dg-(r\dg
r\rho + \rho r\dg r)/2$. This equation represents decoherence of
the atomic system due to photon number fluctuations in the cavity
field, with the result of increased noise in the spin components
normal to $J_{z}$.

The effect of \erf{Hint} on the cavity field is a phase shift
proportional to $J_z$, and thus the output Homodyne photocurrent
is given by \cite{WisMil94} \beq
I_{c}(t)=2\sqrt{M}\an{J_{z}}_{c}+\zeta(t), \label{current} \eeq
where $\zeta(t)$ is a white-noise term satisfying ${\rm
E}[\zeta(t)\zeta(t')]=\delta(t-t')$ and ${\rm E}$ is the ensemble
average. The conditional master equation for the atomic system is
then \cite{WisMil94} \beq d\rho_{c} = dtM{\cal D}[J_{z}]\rho_{c} +
\sqrt{M}dW(t){\cal H}[J_{z}]\rho_{c}, \label{rhoc} \eeq where
$dW(t)=\zeta(t)dt$ is an infinitesimal Wiener increment and ${\cal
H}[r]\rho\equiv r\rho+\rho r\dg-{\rm Tr}[(r+r\dg)\rho]\rho$.

\begin{figure}
\includegraphics[width=0.32\textwidth]{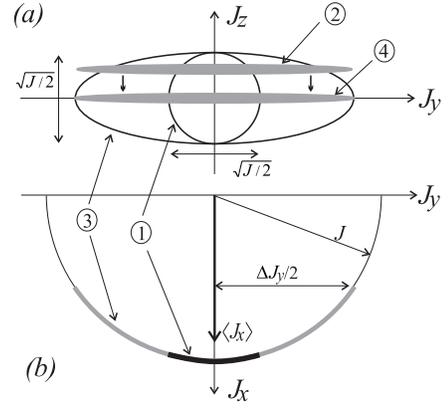}
\vspace{-0.3cm} \caption{\label{states} Schematic $(a)$ $y-z$ and
$(b)$ $x-y$ projections of the quasiprobability distributions for
the spin state. The spin states are represented by ellipses on a
sphere of radius $J$. The initial CSS, spin polarized in the $x$
direction, is given by state 1. State 2 is one particular
conditioned spin state after a measurement of $J_{z}$, while state
3 is the corresponding ensemble average state. The effect of the
feedback is shown by state 4. A rotation about the $y$ axis shifts
the conditioned state 2 back to $\an{J_{z}}_{c}=0$. The ensemble
average of these conditioned states will then be similar to state
4.}
\end{figure}

The effects of this evolution on the initial CSS are a decrease in
the variance of $J_{z}$ with corresponding increases for $J_{y}$
and $J_{x}$ (i.e., spin squeezing), as well as a stochastic shift
of the mean $J_{z}$ away from its initial value of zero. This
shift, indicated by state 2 of Fig.~\ref{states}, is equal to \beq
d\an{J_{z}}_{c}=2\sqrt{M} dW(t) (\Delta J_{z})^{2}_{c} \nn \\
\approx 2\sqrt{M} I_{c}(t)dt \an{J_{z}^{2}}_{c}.
\label{dJzmeas}
\eeq
Here the approximation assumes that $\an{J_{z}}_{c}=0$, which
will be relevant when feedback is in place.

The average or unconditioned evolution, \erf{MEmeas}, is simply
recovered by taking the ensemble average of all possible
conditioned states, i.e., $\rho(t)={\rm E}[\rho_{c}(t)]$. This
leads to a spin state with $J_{z}$ variance equal to $J/2$. In
other words, the unmonitored measurement does not affect $J_{z}$
and the squeezed character of the individual conditioned states is
lost, indicated by state 3 in Fig.~\ref{states}.

To retain the reduced fluctuations of $J_{z}$ in the average
evolution, we employ a feedback mechanism that uses the
measurement record to continuously drive the system into the
$\an{J_{z}}_{c}=0$ squeezed state. The idea is to cancel the
stochastic shift of $\an{J_{z}}_{c}$ due to the measurement. This
simply requires a rotation of the mean spin about the $y$ axis
equal and opposite to that caused by \erf{dJzmeas}, as illustrated
in Fig.~\ref{states}. The feedback Hamiltonian must therefore take
the form \beq H_{\rm fb}(t)=\lambda(t)I_{c}(t)J_{y}/\sqrt{M},
\label{Hfb} \eeq where $\lambda(t)$ is a time-varying feedback
strength. This feedback driving can be implemented by modulating
an applied rf magnetic field \cite{Sangetal01}, as shown in
Fig.~\ref{expt}.

Following again the methods of Ref.~\cite{WisMil94} to find the
total stochastic master equation, we can calculate the conditioned
shift of the mean $J_{z}$ due to the feedback. Again using the
assumption that $\an{J_{z}}_{c}=0$ we have \beq d\an{J_{z}}_{\rm
fb}\approx-\lambda(t) I_{c}(t)dt \an{J_{x}}_{c}/\sqrt{M}.
\label{dJzfeed} \eeq Since the idea is to produce
$\an{J_{z}}_{c}=0$ via the feedback, the approximations above and
in \erf{dJzmeas} apply and we find that the required feedback
strength is \beq \lambda(t)=2M\an{J_{z}^{2}}_{c}/\an{J_{x}}_{c}.
\label{lambdac} \eeq This type of feedback control is essentially
a form of state-estimation-based feedback \cite{Dohetal00}.
Although \erf{Hfb} looks like direct current feedback, the
strength of this feedback (\ref{lambdac}) is determined by
conditioned state expectation values. $I_{c}(t)$ only appears
directly in $H_{\rm fb}$ due to the assumption that the feedback
works and so $\an{J_{z}}_{c}=0$.

Being dependent on conditioned expectation values, which are
computationally very expensive, \erf{lambdac} is not practical in
an experimental sense. What is required is a predetermined series
of data points or ideally an equation for $\lambda(t)$, like in
Fig.~\ref{expt}. To find a suitable expression we begin by
assuming the feedback is successful and replace the conditioned
averages by ensemble averages, \beq \lambda(t)\simeq
2M\an{J_{z}^{2}}/\an{J_{x}}. \label{lambda} \eeq This
approximation will be valid if the unconditioned state has high
purity since then it must comprise of nearly identical highly pure
conditioned states.

The evaluation of both the purity (${\rm Tr}[\rho^{2}]$) and the
averages in \erf{lambda} ($\an{A}\equiv{\rm Tr}[A\rho]$) requires
the unconditioned master equation (ME) \cite{WisMil94} \beq
\dot{\rho}=M{\cal D}[J_{z}]\rho-i\lambda(t)[J_{y},J_{z}\rho+\rho
J_{z}] +\frac{\lambda(t)^{2}}{M}{\cal D}[J_{y}]\rho. \label{ME1}
\eeq The terms in this equation describe, respectively, the noise
due to the measurement back-action, the feedback optical driving,
and the noise introduced by the feedback. The state determined by
\erf{ME1}, with $\lambda(t)$ given by \erf{lambda}, has a purity
very close to one (see below). Since the state is very close to a
pure state, we are justified in applying the approximation of
\erf{lambda}.

Note that \erf{ME1} describes the exact unconditioned evolution of
the atomic system where the feedback strength is arbitrarily
defined by $\lambda(t)$. Equation (\ref{lambda}) thus describes
one particular feedback scheme, however, it can only (easily) be
evaluated numerically. To find an approximate analytical
expression we look at when $\an{\mathbf{J}}\sim{\cal O}(J)$, for
which the atomic sample remains near the minimum uncertainty state
$\an{J_{y}^{2}}\an{J_{z}^{2}}\approx J^{2}/4$. This is equivalent
to a linear approximation represented by replacing $J_{x}$ with
$J$ in the commutator $[J_{y},J_{z}]=iJ_{x}$, which allows us to
calculate $\an{J_{y}^{2}}$ directly from the ME (and hence
$\an{J_{z}^{2}}$). The decrease of $\an{J_{x}}$ from $J$ is then
related to the increase of $\an{J_{y}^{2}}$ from $J/2$ [see
Fig.~\ref{states}$(b)$] due to the measurement back action. Using
these approximations we obtain \beq \lambda(t)\approx
Me^{Mt/2}(1+2JMt)^{-1}. \label{lamapp} \eeq

We can analytically approximate the degree of squeezing produced
by the particular feedback scheme represented by \erf{lamapp}. For
our model \erf{xisq} becomes \beq
\xi^{2}_{z}=\frac{N\an{J_{z}^{2}}}{\an{J_{x}}^{2}}
\simeq\frac{\lambda(t)J}{M\an{J_{x}}} \approx e^{Mt}(1+2JMt)^{-1}.
\label{sqzapp} \eeq This leads to a minimum at $t_{*}\approx 1/M$
of \beq \xi^{2}_{\rm min}\approx e/2J,~~J\gg1. \label{sqzmin} \eeq
Thus, the minimum attainable squeezing parameter asymptotically
approaches an inverse dependence on the sample size, i.e., the
Heisenberg limit \cite{AndLuk01}.

\begin{figure*}
\includegraphics[width=0.64\textwidth]{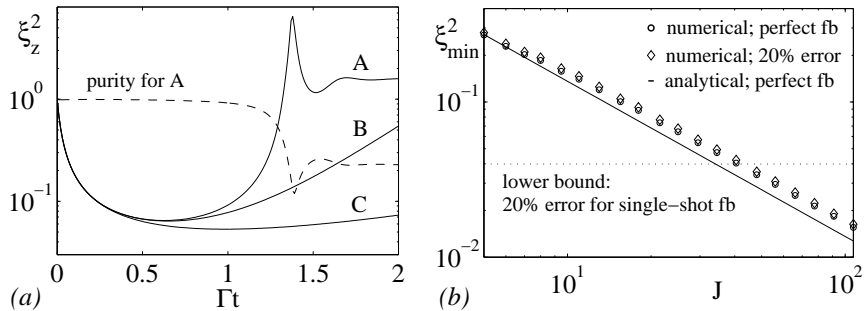}
\vspace{-0.4cm} \caption{\label{xifigs} $(a)$ Time dependence of
the purity $={\rm Tr}[\rho^{2}]$ (dashed curve), and the squeezing
parameter $\xi^{2}_{z}$ (curves A, B, and C). The purity and curve
A are the results for a feedback scheme defined by \erf{lambda}.
For comparison, \erf{sqzexact} is also simulated for the scheme
defined by \erf{lamapp} and the result is curve B. Finally, curve
C is the approximate analytical expression for squeezing parameter
given by \erf{sqzapp}. In all cases $J=25$. $(b)$ $J$ dependence
of the squeezing parameter minimum, $\xi^{2}_{\rm min}$. Plotted
are the results of numerical solutions of the ME, and hence
\erf{sqzexact}, with $\lambda(t)$ given by \erf{lamapp} (circles)
and $\lambda(t)\times120\%$ (diamonds). These values approach
$1.665/J$ and $1.744/J$, respectively. The dotted line is the
lower bound for single-shot feedback also with a $20\%$ error,
while the analytical result for perfect continuous feedback,
\erf{sqzmin}, is the solid line.}
\end{figure*}

The approximations leading up to Eqs.~(\ref{lambda}) and
(\ref{lamapp}) can be justified by numerically solving the ME
(\ref{ME1}) for a given $\lambda(t)$, for example, using the {\sc
Matlab} quantum optics toolbox \cite{Tan99}. The approximation of
\erf{lambda} can be tested by calculating the purity for $\rho$
described by the ME with this particular feedback. The expectation
values in $\lambda(t)$ are found by iteratively solving the ME
[updating $\lambda(t)$ each time step], and thus we also have \beq
\xi^{2}_{z}=2J\an{J_{z}^{2}}/\an{J_{x}}^{2}. \label{sqzexact} \eeq
The results of this simulation are shown in
Fig.~\ref{xifigs}$(a)$, where the purity is given by the dotted
curve and $\xi^{2}_{z}$ is curve A. Clearly, the purity remains
near unity for times of interest. This implies that the
measurement and feedback scheme has worked to produce nearly
identical, nearly pure, conditioned states [for times ${\cal
O}(M^{-1})$], and we are therefore justified in using
\erf{lambda}.

The further approximations to obtain the analytical \erf{lamapp}
are also good, as shown by curve B in Fig.~\ref{xifigs}$(a)$,
where the fit to curve A for the early evolution and minimum is
nearly perfect. We are not interested in later times since the
idea is to cease the measurement and feedback when the minimum
$\xi^{2}_{z}$ is reached. The analytical expression for the
squeezing parameter, \erf{sqzapp}, is also plotted as curve C in
Fig.~\ref{xifigs}$(a)$. Although the minimum is not a perfect fit
to the exact numerical results, it has the correct order of
magnitude and so we expect the $J^{-1}$ scaling of \erf{sqzmin} to
be correct.

The scaling of $\xi^{2}_{\rm min}$ is obtained numerically from
solutions of the ME [and thus \erf{sqzexact}] with feedback
described by \erf{lamapp}. This $\lambda(t)$, shown above to be a
good approximation, is the suitable form for experimental
realization. The numerical results, along with the analytical
expression (\ref{sqzmin}), are plotted in Fig.~\ref{xifigs}$(b)$
which clearly verifies the $J^{-1}$ dependence. The analytical
coefficient ($e/2$) represents an error of $\sim18\%$ compared to
the numerical fit, and the optimum time ($t_{*}$) is also slightly
out as shown by Fig.~\ref{xifigs}$(a)$. Nevertheless, these errors
only apply to scaling coefficients, not to the scalings
themselves.

Experimentally, the limit to squeezing will be dominated by
spontaneous losses due to absorption of QND probe light. The rate
of this loss is $N\gamma\Omega^2/4\Delta^2$, where $\gamma$ is the
spontaneous emission rate and $\Omega=g|\beta|$. To reach the
Heisenberg limit (requiring a time $t_{*}=1/M$) we want the total
loss to be negligible, i.e., we require \beq
N\gamma\frac{\chi|\beta|^2}{\Delta}\frac{\kappa}{4\chi^2|\beta|^2}\sim1
~~\Rightarrow~~ g^2\sim N\kappa\gamma, \label{cavreq}\eeq which is
the very strong coupling regime of cavity QED. Note we also
required $\kappa\gg\chi|\beta|\sqrt{N}$ for the adiabatic
elimination of the cavity field, and to satisfy both we thus
require $\Delta\gg\gamma N^{3/2}$. It is not surprising that
\erf{cavreq} is the same requirement as for Andr\'{e} and Lukin's
model \cite{AndLuk01} implementing the countertwisting Hamiltonian
\cite{KitUed93}, since writing \erf{ME1} in Lindblad form reveals
such a term. Similarly, the condition for achieving some
squeezing, i.e., $\xi^2<1$, will be $g^{2}N > \kappa\gamma$.
Further, we have calculated that a free space model [also given by
\erf{ME1} but with $M$ equal to $N$ in Ref.~\cite{WisTho01}] will
also produce some squeezing, although the Heisenberg limit cannot
be reached since by $t_{*}=1/M$ all atoms will be lost from the
sample.

Figure~\ref{xifigs}$(b)$ also indicates that our continuous scheme
is very robust to any experimental errors in the feedback
strength, as opposed to a single-shot method. The latter approach
consists of a single (integrated) measurement pulse (see e.g.,
Ref.~\cite{Duanetal00}), followed by a single feedback pulse. If
there is a relative error of $\epsilon$ in the feedback strength,
this will induce an error term $(\Delta J_{z}^{\rm err})^{2}\sim
\epsilon^{2}J/2$, which will dominate the total variance for
$J\gg1$. Thus $\xi^2_{\rm min}$ will have a lower bound of
$\epsilon^{2}$, and will never be better than $J^{-1}$. On the
other hand, as shown in Fig.~\ref{xifigs}$(b)$, a large ($20\%$)
error in $\lambda(t)$ for continuous feedback does not affect the
$J^{-1}$ scaling. We have also found this theoretical scaling to
be unaffected by inefficient measurements. Finite feedback delay
time will also have a limited effect as long as it is faster than
$(JM)^{-1}$.

This Rapid Communication has presented a scheme for producing a
spin squeezed atomic sample via QND measurement and feedback. The
advantage over previous QND schemes \cite{KuzBigMan98,KuzManBig00}
is that it provides unconditional, or deterministically
reproducible, squeezing. For very strong cavity coupling, the
theoretical squeezing approaches the Heisenberg limit
$\xi^2\sim1/N$, while some squeezing will be produced at weaker
coupling and even in free space (thus presenting a simple
experimental test for quantum feedback). This indicates a stronger
squeezing mechanism than collisional interactions in a
Bose-Einstein condensate where the scaling is $N^{-2/3}$
\cite{Soretal01,collBEC}. Furthermore, by ceasing the measurement
when this minimum is reached, the maximally squeezed state could
be maintained indefinitely.

L.K.T and H.M.W wish to acknowledge inspiring discussions with
Klaus M\o lmer and Eugene Polzik.

\end{document}